\documentstyle[preprint,aps]{revtex}
\tighten
\include{epsf}
\newcommand{\bel}[1]{\begin{equation}\label{#1}}  
\newcommand{\bal}[1]{\begin{eqnarray}\label{#1}}
\newcommand{\be}{\begin{equation}}
\newcommand{\ee}{\end{equation}}
\newcommand{\ba}{\begin{eqnarray}}
\newcommand{\ea}{\end{eqnarray}}

\newcommand{\del}{\partial}
\newcommand{\eqn}[1]{(\ref{#1})}
\newcommand{\mean}[1]{\langle #1 \rangle}

\newcommand {\bp}{\overline \phi}
\newcommand {\wh}{\widehat}

\newcommand {\wl}{\widehat \lambda}
 \makeatletter                    %allow @ in command names
\@addtoreset{equation}{section}  %reset equation count in each section
\title{ Lattice study of the kink soliton and the zero-mode problem for 
 $\phi^4$ in two dimensions}
\author{A. Ardekani and A. G. Williams\\
Department of Physics and Mathematical Physics and\\
Special Research Center for the Subatomic Structure of Matter,\\
University of Adelaide, SA 5005, Australia}
\begin{document}
\preprint{\parbox[t]{50mm}{\flushright{Preprint: ADP-98-69/T336\\ 
  hep-lat/9811002}}}
\maketitle
\begin{abstract}
\normalsize
\noindent

We study the $\lambda\phi^4_{1+1}$ kink soliton and the zero-mode
 contribution to the kink soliton mass in regions beyond the
 semiclassical regime. The calculations are done in the non-trivial
 scaling region and where appropriate the results are compared with the continuum, semiclassical
 values. We show, as a function of parameter space, where the zero-mode 
 contributions become significant.

\end{abstract}
\vspace{3cm}
\noindent\rule[0mm]{3cm}{0.2mm}
\vspace{1cm}\\
$^\ast$ {\tt aardekan@physics.adelaide.edu.au}\\
$^\ddagger$ {\tt awilliam@physics.adelaide.edu.au}\\
%\end{titlepage}
\newpage
%\chapter {Introduction}
Solitons are non-dispersive localised packets of energy moving uniformly,
and resembling extended particles. The elementary particles in nature are also
localised packets of energy, being described by some quantum field
 theory. 
Because of these features, the soliton might appear as the ideal mathematical 
structure for the description of a particle. When it was realised that many nonlinear field theories used to describe elementary particles,
also had soliton solutions and that these solutions might correspond to particle type 
excitations, the development of methods for soliton quantisation became
 important. The quantisation of solitons is usually done by performing an expansion 
in powers of $\hbar$ (loop expansion) such that the classical soliton solution 
appears as the term of leading order in the expansion and terms of higher order represent the quantum effects. 
 In the mid 1970's there appeared a number of works~\cite{dashen1,ftd,dashen2,dashen3}
which developed the semiclassical expansion in the quantum field theory.
In this period, there were schemes being constructed to quantise these solitons.
The correspondence between classical soliton solutions 
and extended-particle states of the quantised theory were established~\cite {dashen1,dashen2,campleo}. Collective coordinate methods were used to
 deal with the so-called ``zero-mode problem''~\cite {gerv,gerv2,fadkor,rajwb}. This problem is a
 manifestation of the translational invariance of the theory, broken by the
 introduction of the soliton. Field oscillations around this
 classical solution contain zero frequency modes, describing displacements 
 of the soliton. 

Here we study the mass of the simplest topological soliton, that is the
 $\lambda\phi_{1+1}^4$ kink, using lattice Monte Carlo techniques.  The dynamics of this model are governed by a Euclidean Lagrangian density 
\bel {lag}
{\cal L} = {1\over 2} (\del_{\mu}\phi)^2- {\mu^2\over 2}\phi^2+{\lambda\over4}\phi^4 ,
\ee
where $\mu$ is a bare parameter and $\lambda$ is the bare coupling constant. For a free scalar field theory $\lambda \rightarrow 0$ and $-\mu^2\rightarrow m^2$, where $m$ would then be the bare mass of the $\phi$.  In the classical
theory $\mu^2>0$ corresponds to the onset of spontaneous symmetry
breaking.
There are two types of static solutions, both being static solutions, to the equation of motion; the trivial solutions
\be
\phi_0=\pm {\mu\over \sqrt\lambda}\equiv f,
\ee
and the topological solutions
\be
\phi_{\rm k,ak}= \pm {\mu\over \sqrt{\lambda}} \tanh[{\mu\over \sqrt 2}x],
\ee
where $\rm k$ and $\rm ak$ label the topological solutions with the kink and antikink correspond to the positive and negative sign, respectively. The semiclassical regime corresponds to large $f$.

The kink provides an example of a particle with an internal structure,
 distributed over a finite volume rather than concentrated at one point. It
possesses a nonzero,  conserved quantity called the topological charge $Q$ which is defined as
$$
Q=  [\left.\phi(x)\right|_{\infty}-\left.\phi(x)\right|_{-\infty}]
$$
and which leads to stability of the kink solution.
The classical kink mass $M_{\rm cl}$, is defined to be the energy of the static soliton and is given by
\be
 M_{ \rm cl}={2\sqrt{2} \over 3} {\mu^3\over \lambda}.
\ee
 The vacuum and kink solutions can be quantised by path integral
 quantisation or by construction of a tower of approximate harmonic oscillator 
 states around the classical solution $\bp$, where either $\bp= \phi_0,\phi_{\rm k}$ {\rm or} $\phi_{\rm ak}$. In a finite box with the
 length $L$ the soliton mass with a one loop quantum correction becomes~\cite{rajbook}:
\bel {diffe}
M_{\rm sol}=E_{\rm kink}-E_{\rm vac} =M_{\rm cl}+\sum_n{1 \over 2}\omega_n^{2}- 
\sum_n{1 \over 2}\xi_n^{2}+ {\cal{O}} (\wl),
\ee
where $\omega_n$ and $\xi_n$ are the eigenvalues of the following equation
\bel {ev}
\left[-{\del^2\over \del x^2}+\left({3\lambda\phi^2-2 \mu^2}\right)_{\bp}\right]
\eta_i (x)=\Theta_i^2 \eta_i,
\ee
with $\bp=\phi_0$ and $\bp=\phi_k \  {\rm or} \ \phi_{\rm ak}$ for $\omega_n$ and $\xi_n$ respectively. An important point to note is 
that, due to translational invariance, one of the $\omega$'s is zero, i.e, $\omega_0=0$.

 As  $L$ is set to infinity then
 the sums are replaced by integrals and one ends up with a logarithmically divergent integral and renormalisation  is required to render the kink mass finite. Then one arrives at the
  mass of the continuum kink with one loop corrections~\cite{dashen2}:
\bel {semi}
M_{\rm sol}= {2\sqrt(2)\over 3\lambda}\mu^3+\mu({1\over 6}\sqrt{3\over 2}-{3\over \pi
\sqrt{2}})+{\cal O}(\lambda).\label {masseq}
\ee
The zero eigenvalue $\omega=0$ is referred to as the zero-mode and has well known physical
 consequences. These modes always exist when one quantises a theory with a translationally
 invariant Lagrangian about a solution that is not translational invariant. The physical consequence of the existence of the zero-mode is the center of mass motion of the kink. In Eq.~\eqn{masseq} the zero-mode 
contribution to the kink mass is omitted.  That effectively means that the
mass of the quantum kink particle is assumed to be the same as the kink energy. The semiclassical treatment of these zero-modes is done in variety of
ways~\cite {dashen1,jkgs,pol,chlee}, however, we will not discuss these further here, since it is not 
necessary for our purposes.
It is also important to mention that the semiclassical quantisation of the discrete lattice version of the lagrangian density given by Eq.~\eqn{lag} in a finite size box is also complicated by 
 the zero-mode problem.
\begin {center}
{\bf Kink on the lattice}
\end {center}
The $\lambda\phi^4$ action on a 2-d lattice can be written as:
\be
S= -\sum_{n,\mu}  \phi_n\phi_{n+\mu}+\sum_n (2-{\wh \mu^2 \over 2})\phi_n^2+
{\wh \lambda \over 4}\phi^4_n
\ee
where we have defined the dimensionless quantities $\widehat \mu\equiv \mu a$ and $\widehat \lambda\equiv \lambda a^2$ with $a$ being the lattice spacing. In addition $n\equiv (n_1,n_2)$ is a $2$-dimensional vector labeling the lattice sites and $\mu$ is a 
unit vector in the temporal or spatial direction (not to be confused with 
 our parameter $\wh \mu^2\equiv -\wh m^2$). We have also denoted the field on the neighboring site of $n$ in the direction of $\mu$ by $\phi_{n,\mu}$.

This model exhibits two phases. In some regions of the phase space $\mean{\phi}=0$ and these are the
symmetric (unbroken symmetry) regions, whereas in other regions spontaneous 
symmetry breaking occurs and $\mean{\phi}\ne 0$. Classically, for positive
 values of $\wh m^2\equiv -\wh \mu^2$ one always has  $\mean{\phi}=0$ and for negative 
 $\wh m^2$ (i.e., positive $\wh \mu^2$), where spontaneous symmetry breaking occurs,
 $\mean{\phi}\ne 0$.  In this regime the second order critical line
 which separates the two phases is the line corresponding to
 $\wh m^2=0$. Beyond the classical limit the phase space structure changes. There are still two phases and there is still a second order phase critical line separating these phases, however, the position of the critical line
changes and due to quantum fluctuations washing out ``shallow'' spontaneous symmetry breaking it
occurs at a finite negative $\wh m^2$ in general.

In order to determine the critical line, we choose several values of $\wh m^2<0$ located in the broken symmetry sector where $\mean{\phi}\neq 0$. For
 each value of $\wh m^2 $, 
$\wh \lambda$ can be increased until $\mean{\phi}=0$ and, thus, the critical parameters, $(\wh m^2_c,\wh \lambda_c)$, can be found. Of course there is no second order phase transition 
on a finite lattice, however, by a second order phase transition here we mean that the 
correlation length is much larger than the lattice dimensions. The
 critical line is shown in Fig.~1. We have also shown in this figure the 
 one loop prediction for the critical line using the light-front formulation~\cite {light}.

The authors of Ref.~\cite{ciria} proposed two methods of calculating the kink mass on 
the lattice. Here we use one of these methods which uses a local parameter
 and is much less susceptible to 
finite size effects. In this method one imposes an anti-periodic spatial
 boundary condition, giving rise to a topological excitation with 
 non-zero topological charge. Since the kink has the lowest energy in the
 topological sector this topological excitation corresponds to the kink. It is shown that for a 
 a fixed $\wh m$, one has~\cite{ciria}: 
\bel {fullmass}
M_{\rm sol}(\wh \beta)=\int^{\wh \beta}_{\beta_c}d\beta'\Omega(\wh \beta')\equiv {1\over T }\int^{\wh \beta}_{\wh\beta_c}d\beta' { {1\over \beta'}}\left[ \mean {S_a}-\mean{S_p}\right],
\ee
where $M_{\rm sol}$ is the soliton mass, $\wh \beta=1/\wh\lambda$,  $\wh \beta_c$ is the inverse of the
 dimensionless critical bare coupling $\wh \lambda_c$,  $T$
 is the length of the lattice in the temporal direction, $\mean {S_p}$ and
 $\mean {S_a}$ are the mean action of the system with a periodic and anti-periodic spatial boundary condition, respectively.

As we mentioned earlier in the semiclassical quantisation one encounters the zero-mode 
problem with its physical consequences being the center of mass motion of the kink. An interesting 
question is whether this problem persists beyond the semiclassical regime. To answer whether the zero-mode problem persists beyond the semiclassical 
regime, one can examine one of the consequences of 
the existence of a zero-mode, that is, the kink displacements. On a lattice with
 anti-periodic boundary condition in the special direction, we set $\wh m^2\equiv -\mu^2=-1$ and $\wl=4$ giving $f=0.5$, which corresponds to a region beyond the semiclassical regime. Then for an arbitrary  time slice we calculated
$\mean {\phi_n}$ for each site for a number of configurations and an average over these configurations was calculated. We have shown the results
 in Fig.~2. As this figure suggests, the movement of the 
kink due to the translational mode still persists even beyond the semiclassical regime. We repeated the same procedure for different time slices and different number 
 of sampled configurations and these results showed the same
 behaviour.

In order to treat the problem, we imposed an additional constraint, that is $\phi(M)=0$ with $M=(x_0,N/2)$ for all
 times $x_0$, fixing the center of the kink to the center of the lattice. For the  same time slices as in the previous case and for a number of constrained configurations, the value of $\mean{\phi_n}$ for each site  along with the classical configuration are included in Fig.~2. The constrained 
 configuration resembles the kink solution with its center located at
 the center of the lattice. 

Both in numerical MC studies and the analytical calculations it is important 
to find the renormalisation group trajectories (RGT). Along
 these curves and close to an infrared (IR) fixed point the physics described by the lattice regularised quantum field 
is constant and only the value of the cut-off (lattice spacing) is changing. This region is called the scaling region.
The $\lambda \phi^4$ theory in two dimension is an interacting theory. That 
is, in addition to having a Gaussian fixed point at $\wh m^2=\wh \lambda=0$
 where the renormalised coupling $\wh \lambda_r$ 
vanishes, it has other infrared fixed points at which $\wh \lambda_r$ is non-zero. The best candidates for these fixed points are along the critical 
 line where a second order phase transition occurs. In this model the only non trivial critical region is along the transition 
line shown in Fig.~1.  In the vicinity of 
the fixed point the vertex functions strongly scale~\cite{zin} and one expects that close to the critical line, there should be segments of phase space where the ratios of dimensionless vertex functions remain almost constant, giving the scaling region.

In our calculations it is  important to find the scaling region corresponding to a
 non-trivial IR fixed point. That is, one should try to find trajectories 
 away from the Gaussian fixed point. On trivial fixed point trajectories,
 even though spontaneous symmetry breaking can still 
occur and hence a kink solution can exist, the vacuum is governed by a free field. We used $R(\wh m_r,\wh \lambda_r)\equiv\wh m_r^2/ \wh \lambda_r$ as a dimensionless parameter for probing the scaling 
region. This parameter $R$, can be calculated accurately using the effective
 potential method~\cite{me}. The scaling region corresponds to regions where $R$ is approximately constant. Our entire calculations are performed 
 within
 the scaling region so that they can be legitimately  compared with the continuum-semiclassical
 predictions for the soliton mass, given by Eq.~\eqn {semi}. 

For a fixed value of $\wh m^2=-1$ we found a range of values of
 $\wh \lambda$ in the broken symmetry sector, that is $0.2<\wh \lambda<0.8$,  for which the values of  $R(\wh m_r,\wh \lambda_r)$ were
 approximately constant, determining a segment of the scaling
 region corresponding to  $\wh m^2=-1$. Then for some values of
 $\wh \lambda$ within this region we calculated $\Delta S=\mean{S_a}-\mean{S_p}$. In calculations of $\mean{S_a}$ we constrained the center of the kink
 to the center of the lattice. We have plotted $\Delta S/T$ versus $\wh \beta$ along with its classical value in Fig.~3. Having known $\Delta S/T$ , Eq.~\eqn {fullmass} can be used to calculate the soliton
 mass. We also calculated the soliton mass with and without imposing a constraint
 on the center of the kink. The comparison of these results with each
 other and with the classical and semiclassical values is shown in Fig.~4. As is evident, in this case the zero-mode contribution is
 within statistical uncertainties. The statistical uncertainties, as
 one expects, increase as one approaches the critical line which makes 
 the detection of  the zero-mode contribution to soliton mass difficult. 
 Also, it is interesting to note that the Monte Carlo results for
 the soliton mass are less than the classical mass but larger than one
loop semiclassical predictions.  

 Next, for $\wh m^2=-2.2$, we repeated the same procedure and calculated the
 soliton mass for a number of bare couplings within the scaling region.
 The calculations were done with and without a constraint on the center
 of the kink and the results and their comparison with the classical and
 semiclassical predictions are shown in Fig.~5. Our results are 
 consistent with Ref.~\cite {ciria}. There are two important 
 observations. First, unlike the previous case the MC calculated soliton
 masses are lower than the semiclassical values which is due to the higher 
 order corrections. The other important point is that, close to the 
 critical line it appears to be clear that there is a positive contribution 
 to the soliton mass due to the zero-mode. 

 Finally, for $\wh m^2=-4$ we calculated the soliton mass, with and
 without a constraint on the configurations. The results are shown in Fig.~6 and again the zero-mode contribution 
 to the soliton mass appears to be positive. 

 All our calculations were done on a $48\times 48$ lattice. As one
 approaches the critical line the correlation length increases and the
 finite size effects become more significant. However, since the Monte Carlo 
 calculation of masses are based on a local parameter $\Omega(\beta)$ the
 finite size effects are smaller than one might, in general, expect~\cite{twist}. In
 our calculations we kept the correlation length below a half of the lattice
 length. 

 In conclusion we would like to mention that in addition to a
 straightforward elimination of the zero-mode, the imposition of a constraint
 on the center of the kink resulted also in more stable configurations and 
 a reduction of the statistical uncertainties on $\mean {S_a}$ and consequently
 on $M_{\rm sol}$. These instabilities are more significant close to the
 critical line where the field fluctuations are larger. In general we found that the statistical uncertainties 
on $\mean {S_a}$ where much larger than $\mean {S_p}$.

\begin{figure}[htb]
\epsfysize=8.5cm
\centering{\ \epsfbox{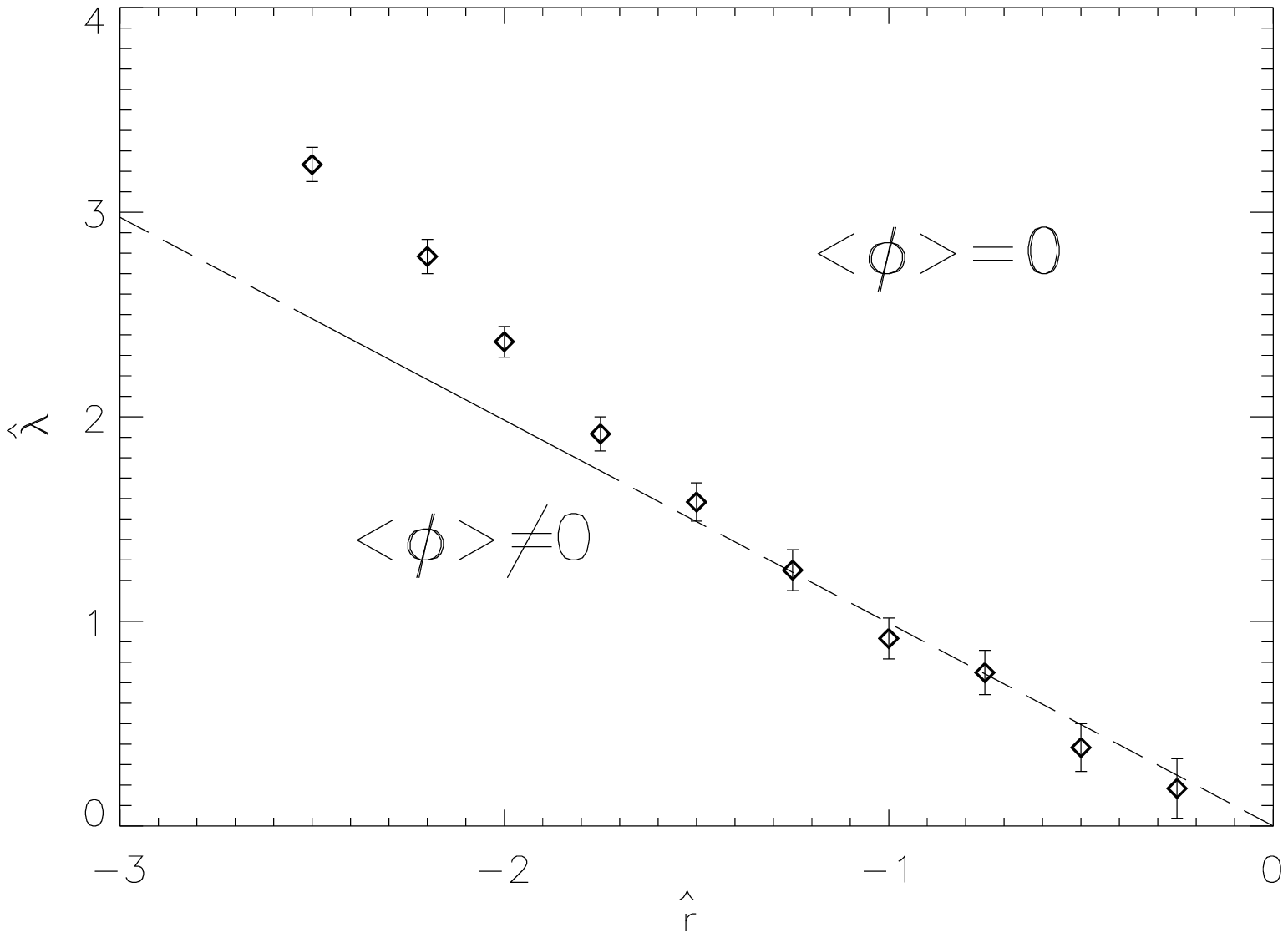}}
\caption{The plot of the transition line between the broken sector and 
unbroken sector using Monte Carlo methods (diamond) and the light-front
 perturbative 
predictions (dashed line). We used the symbol $\wh r \equiv\wh m^2$ in the above figure. }
\end{figure}

\begin{figure}[htb]
\epsfysize=8.5cm
\centering{\ \epsfbox{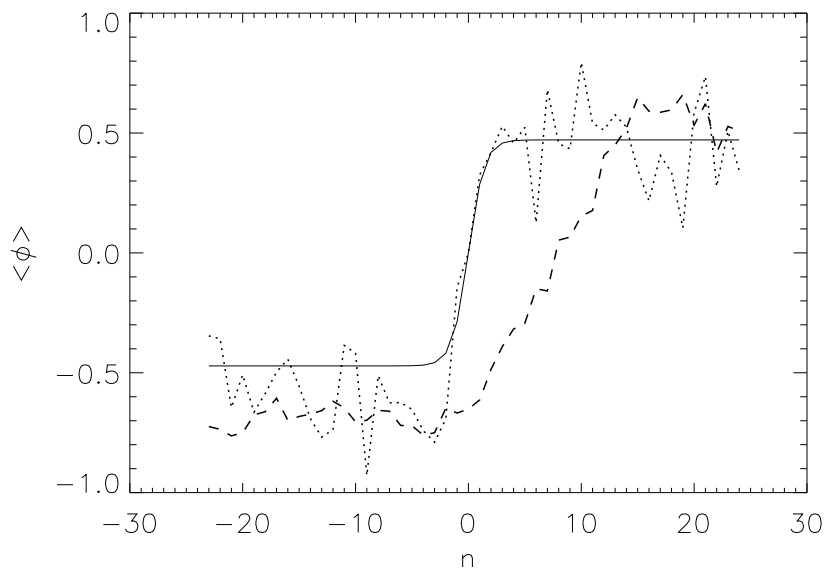}}
\caption{ The comparative plot of averges of fields on a time slice 
with  $f=0.5$
 and $\wh m^2=-1$ for a unconstrained lattice (dashed line) and constrained lattice (dotted line). The constraint fixes the center of the kink to the center of the lattice for all time slices. The solid line is the classical solution.}
\end{figure}

\begin{figure}[htb]
\epsfysize=8.5cm
\centering{\ \epsfbox{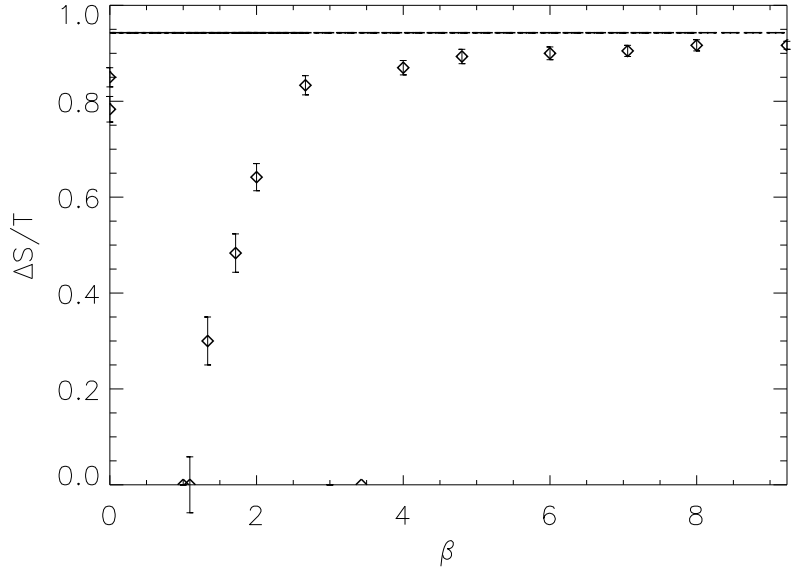}}
\caption{ The plot of $\Delta S/T$ versus $\wh \beta=1/\wh\lambda$ with 
$\wh m^2=-1.0$. The straight horizontal line is the classical value for  $\Delta S/T$ }
\end{figure}

\begin{figure}[htb]
\epsfysize=8.5cm
\centering{\ \epsfbox{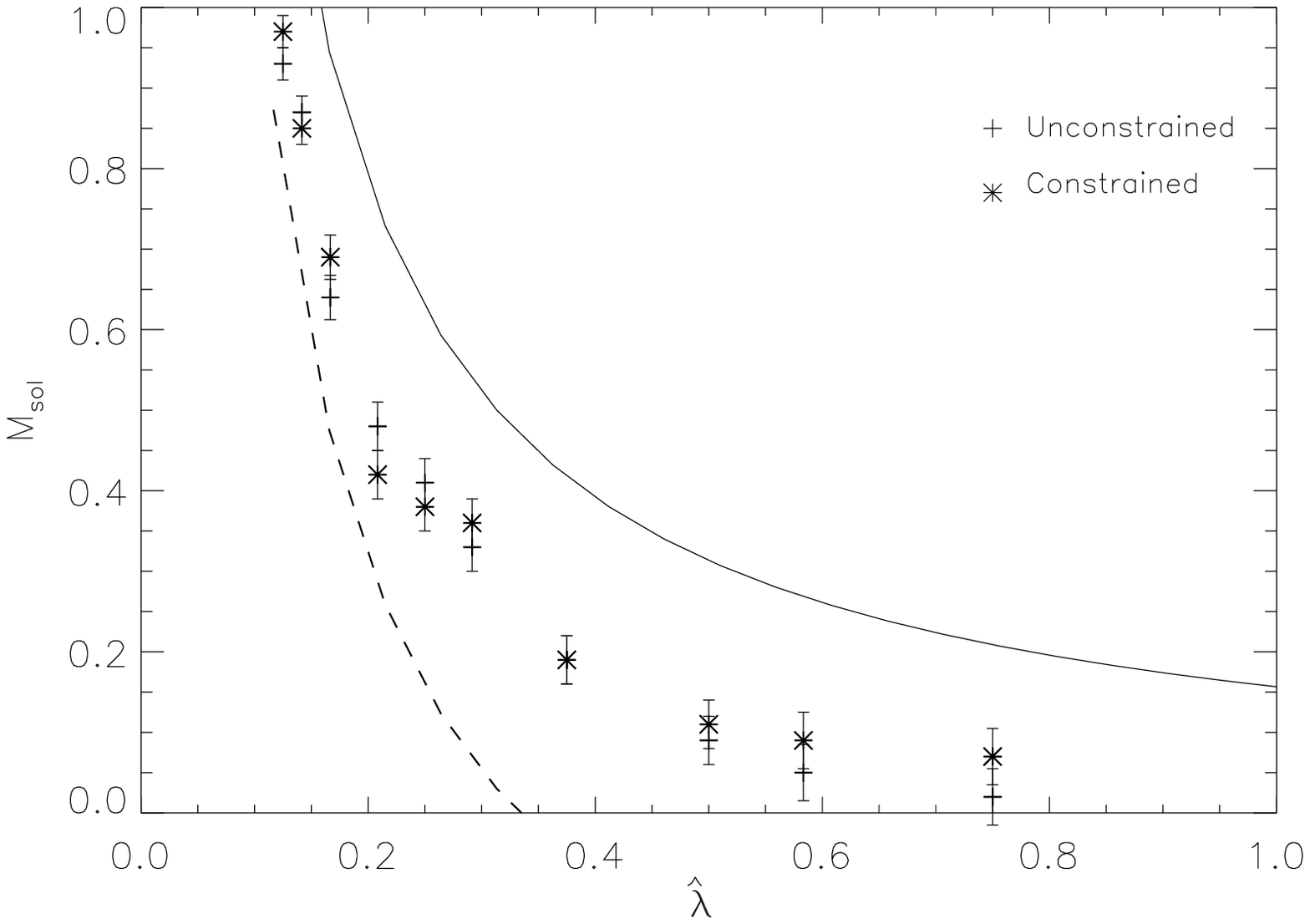}}
\caption{  Plot of soliton mass versus $\beta$ with $\wh m^2=-1$. The Monte Carlo results are compared with the classical (solid line) and semiclassical results (dashed line).}
\end{figure}

\begin{figure}[htb]
\epsfysize=8.5cm
\centering{\ \epsfbox{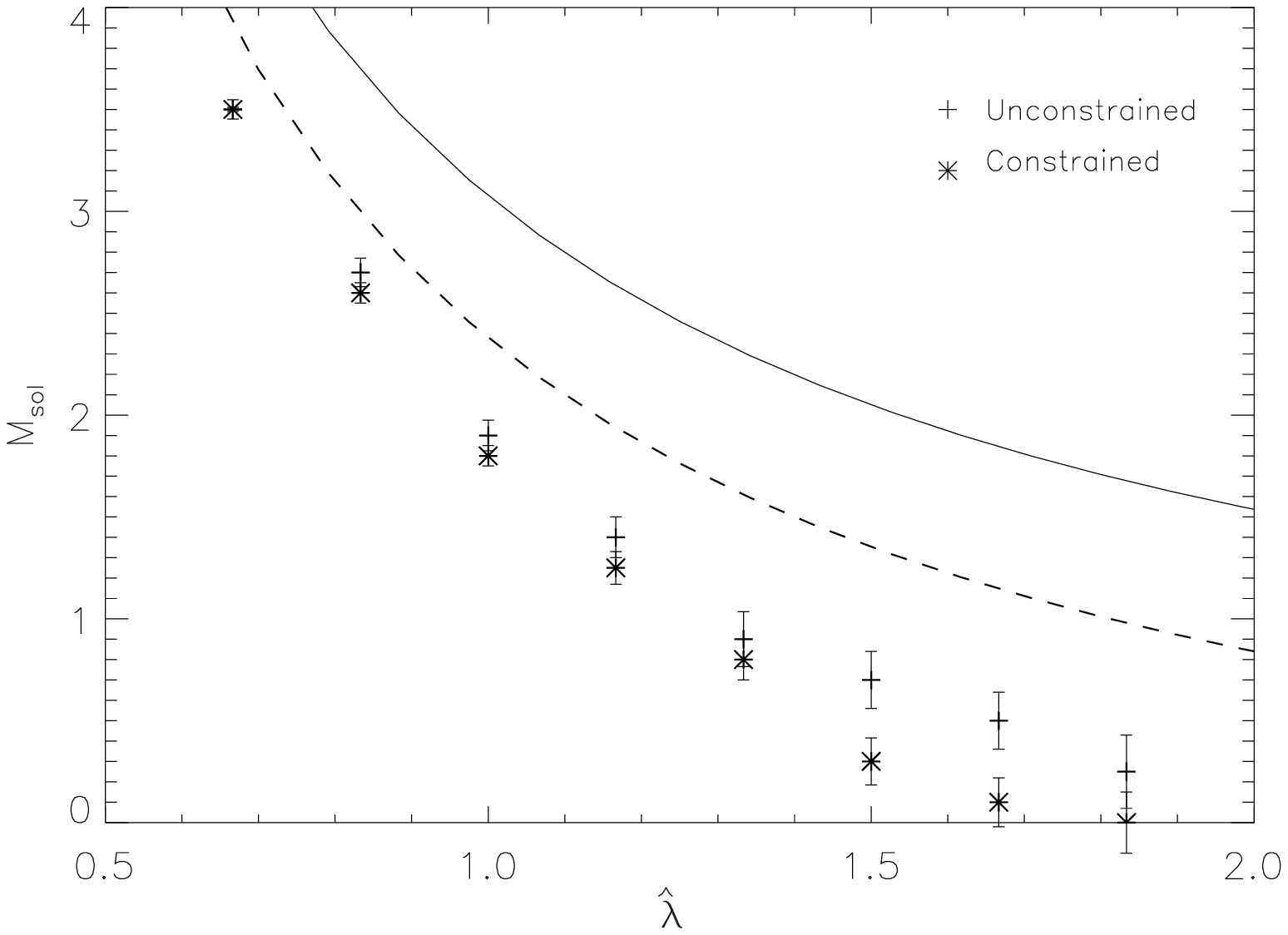}}
\caption{  Plot of soliton mass versus $\beta$ with $\wh m^2=-2.2$. The Monte Carlo results are compared with the classical (solid line) and semiclassical results (dashed line).}
\end{figure}

\begin{figure}[htb]
\epsfysize=8.5cm
\centering{\ \epsfbox{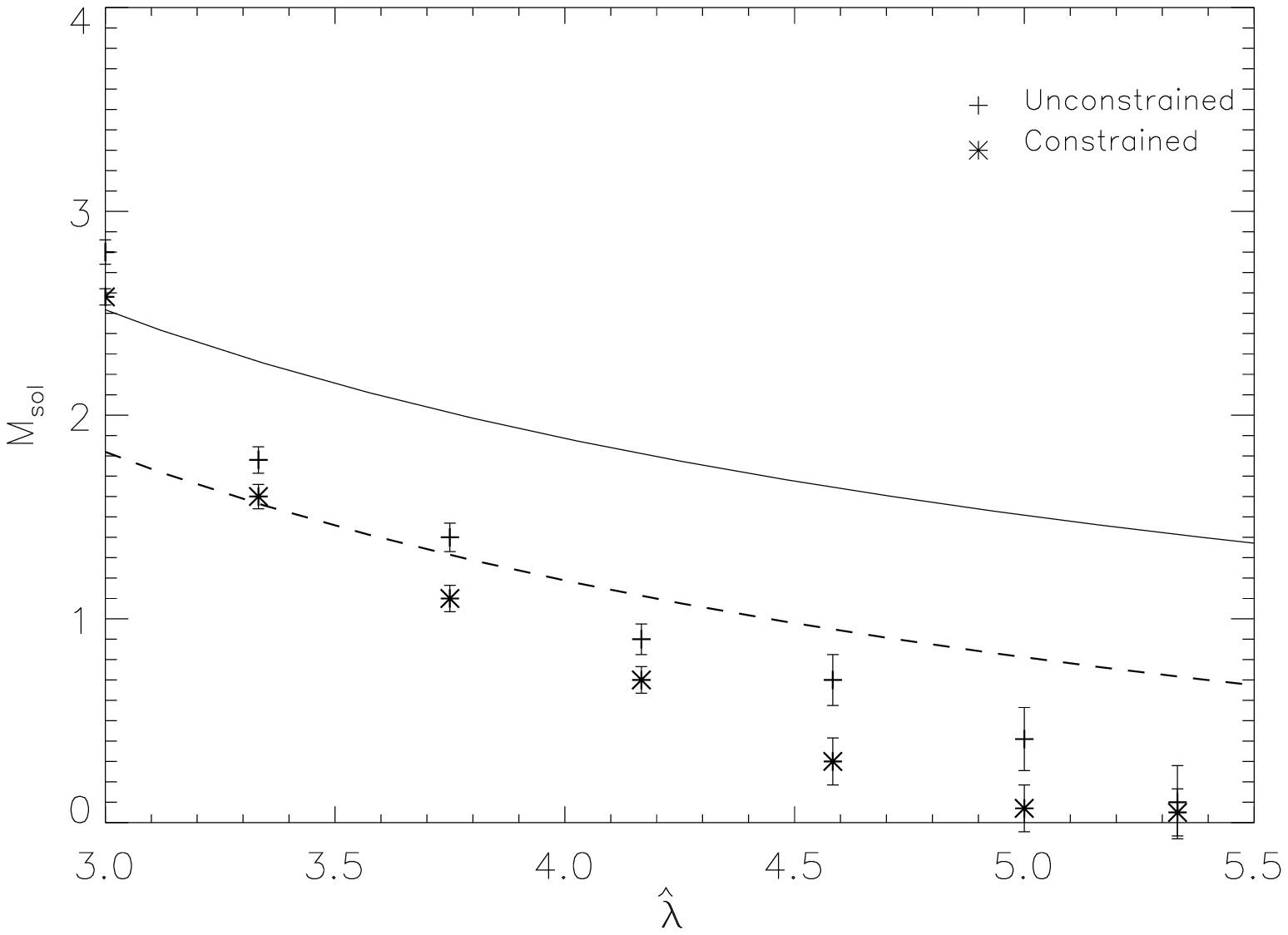}}
\caption{  Plot of soliton mass versus $\beta$ with $\wh m^2=-4$. The Monte Carlo results are compared with the classical (solid line) and semiclassical results (dashed line).}
\end{figure}

\end{document}